\documentstyle[aps,eqsecnum]{revtex}
\renewcommand{\theequation}{\arabic{section}.\arabic{equation}}
\begin{document}
\draft
\title{Nonlinear dynamical systems and classical orthogonal
polynomials}
\author{K. Kowalski}
\address{Department of Theoretical Physics, University
of \L\'od\'z, ul.\ Pomorska 149/153,\\ 90-236 \L\'od\'z,
Poland}
\maketitle
\begin{abstract}
It is demonstrated that nonlinear dynamical systems with
analytic nonlinearities can be brought down to the abstract
Schr\"odinger equation in Hilbert space with boson
Hamiltonian.  The Fourier coefficients of the expansion of
solutions to the Schr\"odinger equation in the particular
occupation number representation are expressed by means of
the classical orthogonal polynomials.  The introduced
formalism amounts a generalization of the classical methods
for linearization of nonlinear differential equations such
as the Carleman embedding technique and Koopman approach.
\end{abstract}
\section{Introduction}
In 1931 Koopman \cite{1} showed that one can associate with
Hamiltonian systems of ordinary differential equations the
Schr\"odinger equation in Hilbert space of square integrable
functions.  The observations of Koopman have become an
important tool in the spectral theory of dynamical systems
\cite{2}.  In 1932 Carleman \cite{3} following ideas
of Poincar\'e and Fredholm demonstrated that nonlinear systems of
ordinary differential equations with polynomial
nonlinearities can be reduced to an infinite system of
linear differential equations.  This approach is nowadays
referred to as the Carleman linearization or Carleman
embedding.  The Carleman approach has been succesfully
applied to the solution of numerous nonlinear problems (see \cite{4}
and references therein).  For example it was used for
calculating Lyapunov exponents \cite{5} and finding first
integrals for the Lorenz system \cite{6}.  In his book
\cite{7} Varadarajan extended the Koopman linearization to
the very general case of the phase space replaced with a
$G$-space.  When $G$=${\bf R}$ and action of the group is
given by the flow the Varadarajan observations can be
regarded as a generalization of the Koopman approach to the
case with non-Hamiltonian systems.  Recently, such a
generalization was rediscovered by Alanson \cite{8} who
reported the possibility of reformulation of dynamical
systems in the Hilbert space of square integrable functions.
In 1982 Steeb \cite{9} demonstrated that the Carleman
embedding matrix can be expressed with the help of Bose
creation and annihilation operators.  Inspired by this
observation the author introduced in 1987 the Hilbert space
approach to nonlinear dynamical systems \cite{10}.  The
formalism is based on the reduction of nonlinear dynamical
systems with analytic nonlinearities to the abstract, linear
Schr\"odinger-like equation in Hilbert space with non-Hermitian boson
Hamiltonian.  The treatment amounts a
far-reaching generalization of the Carleman linearization
technique which corresponds to the particular case with the
occupation number representation for the Schr\"odinger-like
equation.  On the other hand, it works also in the case with
partial differential equations.  The approach has been
developed in a series of papers (see monograph \cite{11} and
references therein) and it has been shown therein to be an
effective tool in the study of both ordinary and partial
differential equations.

As remarked by Carleman \cite{3} (see also \cite{12}) the
simplest polynomial ansatz utilized in his linearization scheme can be
replaced by an orthogonal polynomial.  Indeed, the author
showed in \cite{11} that the Koopman linearization is nothing
but a version of the Carleman technique with the polynomial
linearization ansatz coinciding with a multidimensional
generalization of the Hermite polynomials.  It is worthwhile 
to note that Carleman who discussed the Koopman formalism in 
\cite{3} and wrote down the linearization transformation 
coinciding with that given by Hermite polynomials did not 
recognize such interpretation of the Koopman observations.  
So in both cases of the Carleman and Koopman approach the 
system is linearized via a polynomial ansatz.  We remark that 
the polynomials utilized by Carleman are orthogonal ones only 
in the complex domain.  Now, using the Hilbert space 
formalism introduced by the author one can connect the 
Carleman polynomial linearization ansatz with the 
Schr\"odinger-like equation in Hilbert space.  The reason of 
the nonhermicity of the corresponding Hamiltonian is the fact 
that we deal with complex orthogonal polynomials.  In 
summary, the common feature of the Carleman and Koopman 
approaches is an orthogonal polynomial ansatz which allows to reduce 
the nonlinear dynamical systems to the Schr\"odinger or 
Schr\"odinger-like equation in Hilbert space.  The aim of 
this work is to introduce a general method for linearization 
of nonlinear dynamical systems with analytic nonlinearities 
including the Carleman technique and Koopman approach as a 
special case. Namely, we show that the 
linearization ansatz enabling reduction of dynamical systems 
to the Schr\"odinger equation is generated by an arbitrary 
classical orthogonal polynomial.

The paper is organized as follows.  In section 2 we briefly
recall the Carleman technique and its Hilbert space
generalization.  Section 3 is devoted to a short exposition
of the Koopman linearization.  In section 4 we describe the
method for linearization of nonlinear dynamical systems in
the general case of an arbitrary classical orthogonal
polynomial.  The explicit relations for the concrete
polynomials are discussed in section 5.
\section{Carleman linearization and the Hilbert space
approach to nonlinear dynamical systems}
We first briefly recall the Carleman technique \cite{3}.
Consider the real analytic system
\begin{equation}
\dot {\bf x} = {\bf F}({\bf x}),
\end{equation}
where ${\bf F}:{\bf R}^k\to{\bf R}^k$ and {\bf F} is
analytic in {\bf x}.  Setting
\begin{equation}
\phi_{\bf n} = \prod_{i=1}^{k}x_i^{n_i},
\end{equation}
where ${\bf x}=(x_1,\ldots ,x_k)$ satisfies (2.1) and
$n_i\in{\bf Z}_+$ (the set of nonnegative integers), we
arrive at the linear differential-difference equation such
that
\begin{equation}
\dot\phi_{{\bf n}} = \sum_{{\bf m}\in{\bf Z}_+^k}C_{\bf nm}\phi_{\bf m}.
\end{equation}
Since we can introduce an order in the set ${\bf Z}_+^k$,
therefore (2.3) is equivalent to the infinite 
linear system of ordinary differential equations. In view of
(2.2) the finite system (2.1) is embedded into the infinite
system implied by (2.3).  Indeed, it follows from (2.2) that
the solution {\bf x} to (2.1) is linked to the solution
$\phi_{\bf n}$ of (2.3) by
\begin{equation}
x_i = \phi_{{\bf e}_i},\qquad i=1,\,\ldots \,,\,k,
\end{equation}
where ${\bf e}_i=(0,\ldots ,0,1_i,0,\ldots ,0)$ is the unit
vector of ${\bf R}^k$.  Therefore, the Carleman
linearization is also referred to as the Carleman embedding
technique.

We now outline the Hilbert space approach to nonlinear
dynamical systems \cite{10}.  Consider the system (complex
or real)
\begin{equation}
\dot{\bf z} = {\bf F}({\bf z}),
\end{equation}
where ${\bf F}:{\bf C}^k\to{\bf C}^k$ and {\bf F} is
analytic in {\bf z}.  Let us introduce the vectors in
Hilbert space of the form
\begin{equation}
|{\bf z}\rangle' = 
\exp\left(\frac{1}{2}\sum_{i=1}^{k}|z_i|^2\right)|{\bf z}\rangle,
\end{equation}
where {\bf z} fulfil (2.5) and $|{\bf z}\rangle$ are
normalized coherent states (see appendix B).  On
differentiating both sides of (2.6) with respect to time we
arrive at the following linear, abstract Schr\"odinger-like
equation in Hilbert space satisfied by the vectors $|{\bf
z}\rangle'$:
\begin{equation}
\frac{d}{dt}|{\bf z}\rangle' = M'|{\bf z}\rangle',
\end{equation}
where $M'$ is the boson operator such that
\begin{equation}
M' = \sum_{i=1}^{k}a_i^\dagger F_i({\bf a}).
\end{equation}
Here $a_i^\dagger $, $a_j$, $i=1$, $\ldots\,$,~$k$, are the
standard Bose creation and annihilation operators,
respectively (see appendix A).  Evidently, the solutions to the nonlinear
system (2.5) are linked to the solutions to the linear
equation (2.7) by
\begin{equation}
{\bf a}|{\bf z}\rangle' = {\bf z}|{\bf z}\rangle'.
\end{equation}
It thus appears that the integration of the nonlinear
dynamical system (2.5) can be brought down to the solution
of the linear Schr\"odinger-like equation in Hilbert space
(2.7).

We now discuss the connection of the Hilbert space formalism
with the Carleman linearization.  Writing the abstract
equation (2.7) in the occupation number representation (see 
appendix A) we obtain
\begin{equation}
\dot z_{\bf n} = \sum_{{\bf m}\in{\bf Z}_+^k}M'_{{\bf
nm}}z_{\bf m},
\end{equation}
where $z_{\bf n}=\langle {\bf n}|{\bf z}\rangle'$ and
$M'_{\bf nm}=\langle {\bf n}|M'|{\bf m}\rangle$.  Using
(2.6) and (B.6) we get
\begin{equation}
z_{\bf n} = \prod_{i=1}^{k}\frac{z_i^{n_i}}{\sqrt{n_i!}},
\end{equation}
where ${\bf z}=(z_1,\ldots ,z_k)$ obeys (2.5).  Thus it
turns out that the Carleman embedding technique corresponds
to the particular occupation number representation within
the Hilbert space approach.  We note that the polynomials (2.11)
form the orthonormal complete set in the Fock-Bargmann space
specified by the inner product (B.7) (see (B.9)).  In other words, (2.11)
is simply the normalized version of (2.2).  It should
be noted that in the case with monomials like (2.11) with $k=1$, the
passage to the complex domain is necessary
for obtaining the complete orthonormal set.
Indeed, it can be easily demonstrated that there is no
Hilbert space spanned by real monomials $p_n(x)=c_nx^n$.
However, it might be observed that (2.7) and (2.9) hold in
the real domain as well.
\section{The Koopman linearization}
This section is devoted to a brief exposition of the Koopman
linearization, more precisely its generalization mentioned
in the introduction, with the help of the Hilbert space
approach described in \cite{11}.  Consider the real analytic
system
\begin{equation}
\dot{\bf x} = {\bf F}({\bf x}).
\end{equation}
Let $|{\bf x}\rangle$, where {\bf x} fulfils (3.1), be the
normalized eigenvectors of the position operators (see
appendix C).  Using (C.5) and (C.2) we find that these
vectors satisfy the linear Schr\"odinger-like equation in
Hilbert space of the form
\begin{equation}
\frac{d}{dt}|{\bf x}\rangle = M|{\bf x}\rangle,
\end{equation}
where the operator $M$ is given by
\begin{equation}
M = -i\sum_{i=1}^{k}\hat p_iF_i(\hat {\bf q}),
\end{equation}
where $\hat p_i$, $\hat q_j$, $i=1$, $\ldots\,$,~$k$, are
the momentum and position operators, respectively.  Clearly,
the solutions to the nonlinear system (3.1) are related to
the solutions of the linear equation (3.2) by
\begin{equation}
\hat {\bf q}|{\bf x}\rangle = {\bf x}|{\bf x}\rangle.
\end{equation}
We have thus shown that the nonlinear dynamical system (3.1)
can be cast into the linear Schr\"odinger-like equation
(3.2).  Moreover, an easy calculation based on (3.2) and
(C.3) shows that the vectors defined by
\begin{equation}
|{\bf x}\rangle{}\tilde{} := e^{\frac{1}{2}\int_{0}^{t}{\rm
div}{\bf F}({\bf x})d\tau}|{\bf x}\rangle,
\end{equation}
obey the Schr\"odinger equation of the form
\begin{equation}
i\frac{d}{dt}|{\bf x}\rangle\tilde{} = H|{\bf x}\rangle\tilde{},
\end{equation}
where the Hamiltonian is
\begin{equation}
H = \hbox{$\frac{1}{2}$}\sum_{i=1}^{k}(\hat
p_iF_i(\hat {\bf q}) + F_i(\hat {\bf q})\hat p_i).
\end{equation}
In view of (3.5) the solutions to (3.1) and (3.6) are
related by
\begin{equation}
\hat {\bf q}|{\bf x}\rangle\tilde{} = {\bf x}|{\bf
x}\rangle\tilde{}.
\end{equation}
Thus it turns out that the nonlinear dynamical system (3.1)
can be brought down to the abstract Schr\"odinger equation
(3.6).  Observe that the Hamiltonian $H$ is nothing but the
Hermitian part of the operator $iM$.  One may ask if we
could analogously symmetrize the operator $M'$ given by
(2.8).  The negative answer follows from observations of
Kano \cite{13} who showed that whenever Hamiltonian is not
linear in the Bose creation operators then the coherent
states become unstable, that is the relation (2.9) is
violated during the time evolution.  We finally remark that
the $L^2$ version of the relations (3.5) and (3.6) was
originally introduced by Alanson \cite{8}.
\section{Linearization in Hilbert space and classical 
orthogonal polynomials}
\subsection{Classical orthogonal polynomials}
As we promised in the introduction we now describe a general 
method for reduction of nonlinear dynamical systems to the 
Schr\"odinger-like or Schr\"odinger equation based on the 
linearization ansatz generated by an arbitrary classical 
orthogonal polynomial.  We begin by 
recalling the basic properties of classical orthogonal 
polynomials.  Let $p_n(x)$ be a normalized classical 
orthogonal polynomial, that is we have
\begin{equation}
\int_{a}^{b}dx\,w(x)p_n(x)p_m(x) = \delta_{nm},
\end{equation}
where $w(x)$ is the weight function and $n,\,m\in{\bf Z}_+$.  
The orthogonal polynomials satisfy the recurrence formula
\begin{equation}
p_{n+1}(x) = (A_nx+B_n)p_n(x) - C_np_{n-1}(x),\qquad 
n=0,\,1,\,2,\,\ldots \,,
\end{equation}
where $p_{-1}(x)=0$ and
\begin{equation}
A_n=\frac{q_{n+1}}{q_n},\qquad B_n=A_n(r_{n+1}-r_n),\qquad 
C_n=\frac{A_n}{A_{n-1}}.
\end{equation}
The coefficients $q_n$ and $r_n$ in (4.3) are given by
\begin{eqnarray}
p_n(x) &=& q_nx^n + q'_nx^{n-1}+\ldots ,\nonumber\\
r_n &=&\frac{q'_n}{q_n}.
\end{eqnarray}
We note that the last formula of (4.3) on $C_n$ holds only 
for the normalized orthogonal polynomials.  We now 
specialize to the case with normalized classical orthogonal 
polynomials.  We then have a generalized Rodrigues formula 
\cite{14}
\begin{equation}
p_n(x) = \frac{1}{K_nw(x)}\frac{d^n}{dx^n}[w(x)X^n],
\end{equation}
where $K_n$ are constant and $X$ is a polynomial with 
coefficients independent of $n$, and the differentiation 
formula \cite{14}
\begin{equation}
X\frac{dp_n(x)}{dx} = 
(\alpha_n+\hbox{$\frac{n}{2}$}X''x)p_n(x) + 
\beta_np_{n-1}(x),
\end{equation}
where
\begin{equation}
\alpha_n = 
nX'(0)-\hbox{$\frac{1}{2}$}X''r_n,\qquad 
A_n\beta_n=-C_n[q_1K_1 + 
(n-\hbox{$\frac{1}{2}$})X''].
\end{equation}
\subsection{Reduction to the evolution equation in Hilbert 
space}
We now come to the discussion of the linearization of 
nonlinear dynamical systems in Hilbert space.  As it is 
well-known the orthogonal polynomials form the complete set 
in $L^2_w$ \cite{14}.  Based on this observation we introduce the 
Hermitian operator $\hat x$ with the complete set of 
eigenvectors, such that
\begin{equation}
\hat x|x\rangle = x|x\rangle,
\end{equation}
and
\begin{equation}
\langle n|x\rangle = w(x)^\frac{1}{2}p_n(x),
\end{equation}
where $|n\rangle$, $n\in{\bf Z}_+$, are the basis vectors of the 
occupation number representation (see appendix A) and 
$p_n(x)$ is a normalized classical orthogonal polynomial.  
Evidently, the resolution of the identity for the states 
$|x\rangle$ can be written as
\begin{equation}
\int_{a}^{b} dx\,|x\rangle\langle x| = I.
\end{equation}
On writing the eigenvalue equation (4.8) in the occupation 
number representation and using (4.9), (4.2), (4.3) and 
(A.10) we arrive at the following boson realization of the 
operator $\hat x$:
\begin{mathletters}
\begin{eqnarray}
\hat x &=& a\frac{1}{A_{N-1}\sqrt{N}} + 
\frac{1}{A_{N-1}\sqrt{N}}a^\dagger - \frac{B_N}{A_N}\\
&=& a\frac{1}{A_{N-1}\sqrt{N}} + 
\frac{1}{A_{N-1}\sqrt{N}}a^\dagger + r_N-r_{N+1}, 
\end{eqnarray}
\end{mathletters}
where $N$ is the number operator.  Consider now the 
differential equation
\begin{equation}
\dot x = F(x),
\end{equation}
where $F$ is analytic in $x$.  Our aim is to study the 
dynamics of the time-dependent vectors $|x\rangle$, where 
$x$ satisfies (4.12).  These states will be seen to be 
stable with respect to the time evolution given by (4.12), 
that is (4.8) holds at any time.  Expanding the 
time-dependent vector $|x\rangle$ in the basis of the 
vectors $|n\rangle$, differentiating with respect to time 
and using (4.9), (4.6), (4.2), (4.5) and (A.10) we find that 
the vectors $|x\rangle$ satisfy the Schr\"odinger-like 
equation in Hilbert space of the form
\begin{equation}
\frac{d}{dt}|x\rangle = M|x\rangle,
\end{equation}
where the operator $M$ is
\begin{equation}
M = \hat \pi\frac{F(\hat x)}{X}.
\end{equation}
Here
\begin{eqnarray}
\hat \pi &=& 
\frac{1}{2}\left\{a\frac{1}{A_{N-1}\sqrt{N}}[X''(N-1)+q_1K_1
] - [X''(N-1)+q_1K_1]\frac{1}{A_{N-1}\sqrt{N}}a^\dagger\right.\nonumber\\ 
&&\left.{} + [X''(N-1)+q_1K_1]r_N - [X''N+q_1K_1]r_{N+1} + 
\frac{B_0K_1}{K_0} - X' + 2NX'(0)\right\}.
\end{eqnarray}
In formulas (4.14) and (4.15) we have not designated for 
brevity the dependence of $X$ and $X'$ on $\hat x$.  It 
should also be noted that $X''$ is constant for arbitrary 
classical orthogonal polynomial.  Therefore, $X''$ is a 
$c$-number in (4.15).  Now, it can be verified that the 
following commutation relation holds for arbitrary 
(normalized) classical orthogonal polynomial
\begin{equation}
[\hat x,\hat \pi] = X.
\end{equation}
Therefore, in view of (4.14) we have
\begin{equation}
[\hat x,M] = F(\hat x).
\end{equation}
Hence, with the use of (4.13) we find that (4.16) is 
precisely the condition for stability of the time-dependent 
states $|x\rangle$, i.e.\ (4.8) where $x$ fulfils (4.12), is 
valid at any time.  In other words, the solution to (4.12) 
and the solution of (4.13) are related by (4.8).  It thus 
appears that the solution of the nonlinear differential 
equation (4.12) can be brought down to the solution of the 
linear, evolution, Schr\"odinger-like equation in Hilbert 
space (4.13).  We remark that (4.16) is equivalent to the 
following identities satisfied by an arbitrary normalized 
classical orthogonal polynomial:
\begin{eqnarray}
&&[X''(n-2)+q_1K_1]r_{n-1} - 2[X''(n-1)+q_1K_1]r_n + 
[X''n+q_1K_1]r_{n+1} - 2X'(0) = 0, \\
&&\frac{1}{A_{n-1}^2}[X''(n-\hbox{$\frac{3}{2}$})+q_1K_1] - 
\frac{1}{A_n^2}[X''(n+\hbox{$\frac{1}{2}$})+q_1K_1] = 
X(r_n-r_{n+1}) = X\left(-\frac{B_n}{A_n}\right),
\end{eqnarray}
where $X(r_n-r_{n+1})\equiv X(y)\big\vert_{y=r_n-r_{n+1}}$.  
We finally note that the formula (4.18) is a direct 
consequence of the more general relation such that
\begin{equation}
[X''(n-1)+q_1K_1]r_n - [X''n+q_1K_1]r_{n+1} + 2nX'(0) + 
\frac{B_0K_1}{K_0} = 0.
\end{equation}
\subsection{Reduction to the Schr\"odinger equation}
We now demonstrate that (4.12) can be furthermore reduced to 
the Schr\"odinger equation.  First observe that (4.20) is 
equivalent to the operator formula
\begin{equation}
\hat \pi + \hat \pi^\dagger = -X'.
\end{equation}
Further, owing to (4.16) the Hermitian operator $\hat k$ defined as
\begin{equation}
\hat k = \hbox{$\frac{i}{2}$}(\hat \pi - \hat \pi^\dagger ),
\end{equation}
satisfies the commutation relation
\begin{equation}
[\hat x,\hat k] = iX.
\end{equation}
Using (4.23), (4.22) and (4.21) we find that the Hermitian 
operator $H$ such that
\begin{equation}
H = \frac{1}{2}\left(\hat k\frac{F(\hat x)}{X} + 
\frac{F(\hat x)}{X}\hat k\right),
\end{equation}
is related to the operator $M$ of the form (4.14) by
\begin{equation}
H = i\left(M+\frac{1}{2}\frac{dF}{d\hat x}\right).
\end{equation}
An easy inspection based on (4.13), (4.8) and (4.25) shows 
that the vectors defined by
\begin{equation}
|x\rangle\tilde{} := 
e^{\frac{1}{2}\int_{0}^{t}\frac{dF}{dx}\,d\tau}|x\rangle,
\end{equation}
where $|x\rangle$ fulfils (4.13), satisfy the Schr\"odinger 
equation of the form
\begin{equation}
i\frac{d}{dt}|x\rangle\tilde{} = H|x\rangle\tilde{}.
\end{equation}
By virtue of (4.8) and (4.26) we have
\begin{equation}
\hat x|x\rangle\tilde{} = x|x\rangle\tilde{},
\end{equation}
where $x$ obey (4.12).  Thus it turns out that the solution 
of the nonlinear equation (4.12) can be cast into the 
solution of the Schr\"odinger equation (4.27).  We remark 
that in view of (4.25) the Hamiltonian $H$ coincides with 
the symmetrization of the operator $iM$.
\subsection{Polynomial linearization ansatz}
We note that (4.9) can be regarded as a linearization ansatz 
for (4.12).  Indeed, on writing (4.13) in the occupation 
number representation we arrive at the linear 
differential-difference equation satisfied by $\langle 
n|x\rangle$.  In order to show that the actual treatment can 
be regarded as a generalization of the Carleman 
linearization technique we should reformulate it to deal 
with polynomial linearization ansatz instead of the 
nonpolynomial one given by (4.9).  Consider the vectors
\begin{equation}
|x\rangle' := w(x)^{-\frac{1}{2}}|x\rangle,
\end{equation}
where $|x\rangle$ fulfils (4.13).  Taking into account 
(4.13) we find that these vectors satisfy the following 
evolution equation in Hilbert space:
\begin{equation}
\frac{d}{dt}|x\rangle' = M'|x\rangle',
\end{equation}
where $M'$ is given by
\begin{equation}
M' = \hat \pi'\frac{F(\hat x)}{X},
\end{equation}
where the operator $\hat \pi'$ is
\begin{mathletters}
\begin{eqnarray}
\hat \pi' &=& \hat \pi - \frac{1}{2}\frac{w'}{w}X\\
 &=&\frac{1}{2}\left\{a\frac{1}{A_{N-1}\sqrt{N}}X''(N-1) - 
[X''(N-1)+2q_1K_1]\frac{1}{A_{N-1}\sqrt{N}}a^\dagger\right.\nonumber\\
&&\left.{}-q_1K_1(r_N-r_{N+1}) - \frac{B_0K_1}{K_0}\right\}.
\end{eqnarray}
\end{mathletters}
We remark that in view of (4.32a)
\begin{equation}
[\hat x,\hat \pi'] = X,
\end{equation}
which leads to
\begin{equation}
[\hat x,M'] = F(\hat x).
\end{equation}
Clearly, the solution to (4.12) is linked to the solution 
of (4.30) by
\begin{equation}
\hat x|x\rangle' = x|x\rangle'.
\end{equation}
We have thus shown that the solution to the nonlinear 
equation (4.12) can be reduced to the solution of the linear 
Schr\"odinger-like equation (4.30).  On writing the abstract 
equation (4.30) in the occupation number representation we 
arrive at the following equation:
\begin{equation}
\dot x_n = \sum_{m\in{\bf Z}_+}M'_{nm}x_m,
\end{equation}
where $x_n=\langle n|x\rangle'$ and $M'_{nm}=\langle n|M'
|m\rangle$.  Eqs.\ (4.29) and (4.9) taken together yield
\begin{equation}
x_n = p_n(x).
\end{equation}
Thus it turns out that in the particular occupation number 
representation the presented formalism describes 
linearization of (4.12) by means of the polynomial ansatz 
(4.37) given by an arbitrary classical orthogonal 
polynomial.
\subsection{Linearization of multidimensional nonlinear 
dynamical systems}
We now generalize the actual treatment to the case with 
multidimensional nonlinear dynamical systems.  Having in 
mind the relation (2.2) which is crucial for 
the Carleman linearization, it is 
natural to postulate the following multidimensional 
generalization of classical orthogonal polynomials:
\begin{equation}
p_{\bf n}({\bf x}) = \prod_{i=1}^{k}p_{n_i}(x_i),
\end{equation}
where $p_{n_i}(x_i)$ is a (normalized) classical orthogonal 
polynomial and ${\bf n}\in{\bf Z}_+^k$.  Evidently, $p_{\bf 
n}({\bf x})$ form the orthonormal, complete set in the 
$k$-fold tensor product of $L^2_w$.  We can thus introduce 
the Hermitian operators $\hat x_i$, $i=1$, $\ldots\,$,~$k$, 
with complete set of eigenvectors $|{\bf x}\rangle$, 
${\bf x}\in{\bf R}^k$, satisfying
\begin{eqnarray}
 &&[\hat x_i,\hat x_j] = 0,\qquad i,\,j=1,\,\ldots \,,\,k, \\
 &&\hat {\bf x}|{\bf x}\rangle = {\bf x}|{\bf x}\rangle,\\
&&\langle {\bf n}|{\bf x}\rangle = 
\left(\prod_{i=1}^{k}w(x_i)^{\frac{1}{2}}\right)p_{\bf 
n}({\bf x}),
\end{eqnarray}
where $|{\bf n}\rangle$, ${\bf n}\in{\bf Z}_+^k$, span the 
occupation number representation.  Clearly, the 
resolution of the identity for the states $|{\bf x}\rangle$ 
is
\begin{equation}
\int\limits_{(a,b)^k} d^kx\,|{\bf x}\rangle\langle {\bf x}| = I.
\end{equation}
Furthermore, it can be easily checked (see (A.1) and (A.2)) that the 
$k$-dimensional generalization of (4.11) is
\begin{mathletters}
\begin{eqnarray}
\hat x_i &=& a_i\frac{1}{A_{N_i-1}\sqrt{N_i}} + 
\frac{1}{A_{N_i-1}\sqrt{N_i}}a_i^\dagger - 
\frac{B_{N_i}}{A_{N_i}}\\
&=& a_i\frac{1}{A_{N_i-1}\sqrt{N_i}} + 
\frac{1}{A_{N_i-1}\sqrt{N_i}}a_i^\dagger + 
r_{N_i}-r_{N_i+1},\qquad i=1,\,\ldots \,,\,k. 
\end{eqnarray}
\end{mathletters}
Consider now the nonlinear dynamical system
\begin{equation}
\dot{\bf x} = {\bf F}({\bf x}),
\end{equation}
where ${\bf F}:{\bf R}^k\to{\bf R}^k$ and {\bf F} is 
analytic in {\bf x}.  Proceeding as with (4.12) we find that 
the time-dependent vectors $|{\bf x}\rangle$, where {\bf x} 
fulfils (4.44), obey
\begin{equation}
\frac{d}{dt}|{\bf x}\rangle = M|{\bf x}\rangle.
\end{equation}
The boson operator $M$ is given by
\begin{equation}
M = \sum_{i=1}^{k}\hat \pi_i\frac{F_i(\hat {\bf x})}{X(\hat 
x_i)},
\end{equation}
where
\begin{eqnarray}
\hat \pi_i &=& 
\frac{1}{2}\left\{a_i\frac{1}{A_{N_i-1}\sqrt{N_i}}[X''(N_i-1)+q_1K_1
] - 
[X''(N_i-1)+q_1K_1]\frac{1}{A_{N_i-1}\sqrt{N_i}}a_i^\dagger\right.\nonumber\\ 
&&\left.{} + [X''(N_i-1)+q_1K_1]r_{N_i} - 
[X''N_i+q_1K_1]r_{N_i+1} + 
\frac{B_0K_1}{K_0} - X'(\hat x_i) + 2N_iX'(0)\right\},\\ 
&&i=1,\,\ldots \,,\,k.\nonumber
\end{eqnarray}
From (4.43), (4.47) and (4.16) (see also (A.1) and (A.2)) it 
follows that
\begin{equation}
[\hat x_i,\hat \pi_j] = \delta_{ij}X(\hat x_j),\qquad 
i,\,j=1,\,\ldots \,,\,k.
\end{equation}
Hence
\begin{equation}
[\hat {\bf x},M] = {\bf F}(\hat {\bf x}).
\end{equation}
As with (4.16) we find that (4.48) is the condition for the 
stability of the time-dependent states $|{\bf x}\rangle$.  
Therefore, the eigenvalue equation (4.40) is valid at any 
time.  In other words, the nonlinear dynamical system (4.44) 
can be brought down to the solution of the linear evolution 
equation in Hilbert space (4.45).

It is easy to show using observations of section C that 
(4.44) can be cast into the Schr\"odinger equation.  Namely, 
we introduce the Hermitian operators
\begin{equation}
\hat k_i = \hbox{$\frac{i}{2}$}(\hat \pi_i - \hat 
\pi_i^\dagger ),\qquad i=1,\,\ldots \,,\,k,
\end{equation}
which in view of (4.48) obey
\begin{equation}
[\hat x_r,\hat k_s] = i\delta_{rs}X(\hat x_s),\qquad 
r,\,s=1,\,\ldots \,,\,k.
\end{equation}
Taking into account (4.51), (4.46) and the generalization of 
(4.21) such that
\begin{equation}
\hat \pi_i + \hat \pi_i^\dagger = - \frac{dX(\hat 
x_i)}{d\hat x_i},\qquad i=1,\,\ldots \,,\,k,
\end{equation}
we find that the Hermitian operator $H$ defined as
\begin{equation}
H = \frac{1}{2}\sum_{i=1}^{k}\left(\hat k_i\frac{F_i(\hat {\bf x})}{X(\hat 
x_i)} + 
\frac{F_i(\hat {\bf x})}{X(\hat x_i)}\hat k_i\right),
\end{equation}
is linked to the operator $M$ by
\begin{equation}
H = i(M+\hbox{$\frac{1}{2}$}{\rm div}{\bf F}).
\end{equation}
Therefore, the vectors
\begin{equation}
|{\bf x}\rangle\tilde{} := e^{\frac{1}{2}\int_{0}^{t}{\rm 
div}{\bf F}\,d\tau}|{\bf x}\rangle,
\end{equation}
where $|{\bf x}\rangle$ obey (4.45), satisfy the 
Schr\"odinger equation
\begin{equation}
i\frac{d}{dt}|{\bf x}\rangle\tilde{} = H|{\bf 
x}\rangle\tilde{}.
\end{equation}
Obviously, the following relation holds:
\begin{equation}
\hat {\bf x}|{\bf x}\rangle\tilde{} = {\bf x}|{\bf 
x}\rangle\tilde{},
\end{equation}
where {\bf x} fulfil (4.44).  It thus appears that the 
nonlinear dynamical system (4.44) can be reduced to the 
abstract Schr\"odinger equation (4.56).

We finally discuss the multidimensional generalization of 
the actual treatment in the case with the polynomial 
linearization ansatz (4.37).  The resulting formalism 
generalizes the approach taken up by Carleman.  Let us 
introduce the vectors of the form
\begin{equation}
|{\bf x}\rangle' = 
\left(\prod_{i=1}^{k}w(x_i)^{-\frac{1}{2}}\right)|{\bf 
x}\rangle,
\end{equation}
where $|{\bf x}\rangle$ obey (4.45).  Proceeding as in the 
case with (4.29) we arrive at the following evolution 
equation in Hilbert space satisfied by the vectors (4.58):
\begin{equation}
\frac{d}{dt}|{\bf x}\rangle' = M'|{\bf x}\rangle',
\end{equation}
where $M'$ is
\begin{equation}
M' = \sum_{i=1}^{k}\hat \pi_i'\frac{F_i(\hat {\bf x})}{X(\hat 
x_i)}.
\end{equation}
Here the operators $\hat \pi_i'$, $i=1$, $\ldots\,$,~$k$, 
are given by
\begin{mathletters}
\begin{eqnarray}
\hat \pi_i' &=& \hat \pi_i - \frac{1}{2}\frac{w'(\hat 
x_i)}{w(\hat x_i)}X(\hat x_i)\\
&=&\frac{1}{2}\left\{a_i\frac{1}{A_{N_i-1}\sqrt{N_i}}X''(N_i-1) - 
[X''(N_i-1)+2q_1K_1]\frac{1}{A_{N_i-1}\sqrt{N_i}}a_i^\dagger\right.\nonumber\\
&&\left.{}-q_1K_1(r_{N_i}-r_{N_i+1}) - \frac{B_0K_1}{K_0}\right\}.
\end{eqnarray}
\end{mathletters}
Notice that
\begin{eqnarray}
&&[\hat x_i,\hat \pi_j'] = \delta_{ij}X(\hat x_j),\qquad 
i,\,j=1,\,\ldots \,,\,k,\\
&&[\hat {\bf x},M'] = {\bf F}(\hat {\bf x}).
\end{eqnarray}
Clearly,
\begin{equation}
\hat {\bf x}|{\bf x}\rangle' = {\bf x}|{\bf x}\rangle'.
\end{equation}
Thus it turns out that the nonlinear dynamical system (4.44) 
can be brought down to the linear Schr\"odinger-like 
equation in Hilbert space (4.59).  Writing (4.59) in the 
occupation number representation we get
\begin{equation}
\dot x_{\bf n} = \sum_{{\bf m}\in{\bf Z}_+^k}M'_{\bf 
nm}x_{\bf m},
\end{equation}
where $x_{\bf n}=\langle {\bf n}|{\bf x}\rangle'$ and 
$M'_{\bf nm}=\langle {\bf n}|M'|{\bf m}\rangle$.  Taking 
into account (4.58) and (4.41) we find
\begin{equation}
x_{\bf n} = p_{\bf n}({\bf x}).
\end{equation}
In view of the form of the relations (2.2), (2.3), (4.66), 
(4.38) and (4.65) it is plausible to treat the actual 
formalism as a generalization of the Carleman embedding 
technique to the case with the linearization ansatz given by 
an arbitrary classical orthogonal polynomial.  We recall 
that the orthogonal polynomials (2.13) are complex.  
Nevertheless, as we demonstrate in the next section the 
introduced approach works also in such the case.
\section{Examples}
\subsection{Complex orthogonal polynomials}
In this section we illustrate the general formalism 
described above by the examples of the concrete classical 
orthogonal polynomials.  We begin with the simplest case of 
the complex polynomials (2.13) for $k=1$.  As a matter of fact, these 
polynomials are not considered as classical ones.  
Nevertheless, they satisfy the most important relations 
characteristic for classical orthogonal polynomials such as 
the recurrence, Rodrigues and differentiation formulas.  No wonder that the 
introduced approach covers the case of polynomials (2.13).  
To see this consider the complex polynomials
\begin{equation}
p_n(z) = \frac{z^n}{\sqrt{n!}}.
\end{equation}
Evidently, for these polynomials (see (4.1), (4.2), (4.4) and 
(2.14))
\begin{eqnarray}
 &&w(z,z^*) = e^{-zz^*},\qquad 
A_n=\frac{1}{\sqrt{n+1}},\qquad B_n=0,\qquad C_n=0, \\
 &&q_n=\frac{1}{\sqrt{n!}},\qquad r_n=0.
\end{eqnarray}
We remark that the formula (4.3) on $C_n$ does not hold in 
the case with complex orthogonal polynomials.  Furthermore, 
the Rodrigues formula can be written as
\begin{equation}
p_n(z) = 
\frac{1}{K_nw(z,z^*)}\frac{d^n}{dz^{*n}}[w(z,z^*)X^n],
\end{equation}
where
\begin{equation}
X=1,\qquad K_n=(-1)^n\sqrt{n!}\,.
\end{equation}
It can be easily checked with the use of (5.2), (5.3) and 
(5.5) that the differentiation formula (4.6) is also valid 
by the polynomials (5.1).  Now we can write the complex 
counterpart of relations (4.8) and (4.9) of the form
\begin{eqnarray}
 &&\hat z|z\rangle = z|z\rangle, \\
 &&\langle n|z\rangle = w(z,z^*)^\frac{1}{2}p_n(z).
\end{eqnarray}
Proceeding as with (4.8) and (4.9) we find
\begin{equation}
\hat z = a,
\end{equation}
that is the states $|z\rangle$ satisfying (5.6) are nothing 
but the standard coherent states (see appendix B).  Referring back to the 
earlier remark concerning the coefficient $C_n$ in the 
recurrence formula (4.2) we note that the form of the 
operator $\hat z$ is implied by (5.2) and
\begin{equation}
\hat z = a\frac{1}{A_{N-1}\sqrt{N}} + 
\frac{C_N}{A_N\sqrt{N}}a^\dagger - \frac{B_N}{A_N}.
\end{equation}
This formula is a generalization of (4.11) in the case with 
complex polynomials when the last relation of (4.3) does not 
take place.

Consider the differential equation
\begin{equation}
\dot z = F(z),
\end{equation}
where $F$ is analytic in $z$.  In opposite to the case of 
classical orthogonal polynomials the time-dependent vectors 
$|z\rangle$, where $z$ fulfil (5.10) do not satisfy the 
linear evolution equation in Hilbert space.  Indeed, 
applying the algorithm described in section 4 we get
\begin{equation}
\frac{d}{dt}|z\rangle = [{\rm Re}F(z) + a^\dagger 
F(a)]|z\rangle.
\end{equation}
On the other hand, the vectors defined as
\begin{equation}
|z\rangle' = w(z,z^*)^{-\frac{1}{2}}|z\rangle = 
e^{\frac{1}{2}zz^*}|z\rangle,
\end{equation}
which are the complex counterparts of the vectors (4.29) are 
easily seen to obey
\begin{equation}
\frac{d}{dt}|z\rangle' = M'|z\rangle',
\end{equation}
where the boson operator $M'$ is
\begin{equation}
M' = a^\dagger F(a).
\end{equation}
Notice that since $X=1$ (see (5.5)), the operator (5.14) has 
the structure analogous to (4.31).  Moreover, identifying 
$\hat \pi'$ with $a^\dagger $ we see that (4.33) is valid as 
well.  Furthermore, as an immediate consequence of (5.14) 
and (A.1) we find
\begin{equation}
[a,M'] = F(a).
\end{equation}
As with (4.17) we conclude that the eigenvalue equation
\begin{equation}
a|z\rangle' = z|z\rangle',
\end{equation}
where $z$ satisfies (5.10), holds at any time, that is the 
nonlinear equation (5.10) reduces to the linear 
Schr\"odinger-like equation in Hilbert space (5.13).

The generalization of the linearization algorithm in the 
case with the nonlinear dynamical system (2.5)
is straightforward.  Evidently, the 
counterparts of (4.38), (4.39), (4.40), (4.41) and (4.58) 
can be written as
\begin{mathletters}
\begin{eqnarray}
 &&p_{\bf n}({\bf z}) = \prod_{i=1}^{k}p_{n_i}(z_i) = 
\prod_{i=1}^{k}\frac{z_i}{\sqrt{n_i!}}, \\
 &&[a_i,a_j] = 0,\qquad i,\,j=1,\,\ldots \,,\,k,\\
&&{\bf a}|{\bf z}\rangle = {\bf z}|{\bf z}\rangle,\\
&&\langle {\bf n}|{\bf z}\rangle = 
\left(\prod_{i=1}^{k}w(z_i,z_i^*)^\frac{1}{2}\right)p_{\bf 
n}({\bf 
z})=\exp\left(-\frac{1}{2}\sum_{i=1}^{k}|z_i|^2\right)p_{\bf 
n}({\bf z}),\\
&&|{\bf z}\rangle' = 
\left(\prod_{i=1}^{k}w(z_i,z_i^*)^{-\frac{1}{2}}\right)
|{\bf z}\rangle = 
\exp\left(\frac{1}{2}\sum_{i=1}^{k}|z_i|^2\right)|{\bf z}\rangle.
\end{eqnarray}
\end{mathletters}
Proceeding analogously as in the case of (4.58) we arrive at eq.\
(2.7).  Furthermore, identifying $\hat \pi_i'$ with $a_i^\dagger $ 
we find that the formula (4.60) takes place also in the case 
with complex orthogonal polynomials.  Using (2.8) and (A.1) 
we get the following counterpart of the relation (4.63):
\begin{equation}
[{\bf a},M'] = {\bf F}({\bf a}).
\end{equation}
Clearly, (5.18) leads to the formula (2.9).  Finally, the formula 
(2.11) is a direct consequence of (5.17e).  We have thus 
shown that the particular case with the complex orthogonal 
polynomials (5.17a) corresponds to the Hilbert space 
approach discussed in section 2.
\subsection{Hermite polynomials}
We now examine the case of the Hermite polynomials within 
the actual treatment.  Let $H_n(x)$ be the Hermite 
polynomials.  For these polynomials we have (see (4.1))
\begin{equation}
a=-\infty,\qquad b=\infty,\qquad w(x)=e^{-x^2}.
\end{equation}
Now let $p_n(x)$ be normalized Hermite polynomials, i.e.
\begin{equation}
p_n(x) = 
\pi^{-\frac{1}{4}}2^{-\frac{n}{2}}\frac{1}{\sqrt{n!}}H_n(x).
\end{equation}
The polynomials $p_n$ satisfy the recurrence formula (4.2), 
where
\begin{equation}
A_n=\sqrt{\frac{2}{n+1}},\qquad B_n=0,\qquad 
C_n=\sqrt{\frac{n}{n+1}}.
\end{equation}
These formulas can be recovered from (4.3) with the help of 
the relations
\begin{equation}
q_n=\pi^{-\frac{1}{4}}2^{\frac{n}{2}}\frac{1}{\sqrt{n!}},
\qquad r_n=0.
\end{equation}
The Rodrigues formula for polynomials $p_n$ is given by 
(4.5), where
\begin{equation}
X=1,\qquad 
K_n=\pi^{\frac{1}{4}}(-1)^n2^\frac{n}{2}\sqrt{n!}\,.
\end{equation}
Using (5.21) we find that the operator (4.11) is
\begin{equation}
\hat x = \frac{1}{\sqrt{2}}(a + a^\dagger ),
\end{equation}
that is
\begin{equation}
\hat x = \hat q,
\end{equation}
where $\hat q$ is the position operator (see appendix C).  
Furthermore, taking into account (5.21), (5.22) and (5.23) 
we obtain the following formula on the operator (4.15):
\begin{equation}
\hat \pi = \frac{1}{\sqrt{2}}(a^\dagger - a).
\end{equation}
A look at (5.26) is enough to conclude that
\begin{equation}
\hat \pi = -i\hat p,
\end{equation}
where $\hat p$ is the momentum operator (see appendix C).  
Therefore, the operator given by (4.22) coincides with the 
momentum operator, i.e.
\begin{equation}
\hat k = \hat p.
\end{equation}
We now return to the Schr\"odinger-like equation (4.45).  
Eqs.\ (4.46), (4.43), (4.47), (5.23), (5.25) and (5.27) 
taken together lead to the operator $M$ given by (3.3).
Thus, the nonlinear system (4.44) reduces to the linear 
Schr\"odinger-like equation (4.45), where $M$ is expressed by (3.3).  
Moreover, by virtue of (4.53) and (5.28) the system can be
furthermore brought down to the Schr\"odinger equation 
(4.56), where the Hamiltonian $H$ is given by (3.7).
It thus appears that the particular case 
of the Hermite polynomials within the presented approach 
refers to the Koopman linearization.
\subsection{Jacobi polynomials}
Our purpose now is to discuss the case of the Jacobi 
polynomials within the introduced formalism.  The Jacobi 
polynomials $P_n^{(\alpha,\beta)}(x)$ are specified by
\begin{equation}
a=-1,\qquad b=1,\qquad w(x)=(1-x)^\alpha(1+x)^\beta,
\end{equation}
where $\alpha>-1$ and $\beta>-1$.  Let $p_n$ be the 
normalized Jacobi polynomials, so
\begin{equation}
p_n(x) = 
\left[\frac{(2n+\alpha+\beta+1)\Gamma(n+1)\Gamma(n+\alpha+
\beta+1)}{2^{\alpha+\beta+1}\Gamma(n+\alpha+1)\Gamma(n+\beta
+1)}\right]^\frac{1}{2}P_n^{(\alpha,\beta)}(x).
\end{equation}
The coefficients in the recurrence formula (4.2) are of the 
form
\begin{eqnarray}
A_n &=& 
\frac{2n+\alpha+\beta+2}{2}\sqrt{\frac{(2n+\alpha+\beta+1)(2n+
\alpha+\beta+3)}{(n+1)(n+\alpha+1)(n+\beta+1)(n+\alpha+\beta
+1)}}\,\,,\nonumber\\
B_n &=& 
\frac{\alpha^2-\beta^2}{2(2n+\alpha+\beta)}\sqrt{\frac{(2n+
\alpha+\beta+1)(2n+\alpha+\beta+3)}{(n+1)(n+\alpha+1)(n+
\beta+1)(n+\alpha+\beta+1)}}\,\,,\nonumber\\
C_n &=& 
\frac{2n+\alpha+\beta+2}{2n+\alpha+\beta}\sqrt{\frac{n(n+
\alpha)(n+\beta)(n+\alpha+\beta)(2n+\alpha+\beta+3)}{(n+1)(n
+\alpha+1)(n+\beta+1)(n+\alpha+\beta+1)(2n+\alpha+\beta-1)}}
\,\,.
\end{eqnarray}
Accordingly, the coefficients $q_n$ and $r_n$ (see (4.4)) 
are
\begin{eqnarray}
q_n &=& 
\left[\frac{2n+\alpha+\beta+1}{2^{2n+\alpha+\beta+1}\Gamma(n
+1)\Gamma(n+\alpha+1)\Gamma(n+\beta+1)\Gamma(n+\alpha+\beta+
1)}\right]^\frac{1}{2}\Gamma(2n+\alpha+\beta+1),\nonumber\\
r_n &=& \frac{(\alpha-\beta)n}{2n+\alpha+\beta}.
\end{eqnarray}
The expressions for the polynomial $X$ and the coefficient 
$K_n$ in the Rodrigues formula (4.5) are
\begin{equation}
X = 1-x^2,\qquad
K_n 
= (-1)^n\left[\frac{2^{2n+\alpha+\beta+1}\Gamma(n+1)\Gamma
(n+\alpha+1)\Gamma(n+\beta+1)}{(2n+\alpha+\beta+1)\Gamma(n+
\alpha+\beta+1)}\right]^\frac{1}{2}\,\,.
\end{equation}
Now taking into account (5.31) we find that the formula 
(4.11) takes the form
\begin{eqnarray}
&&\hat x = 
a\,\frac{2}{2N+\alpha+\beta}\sqrt{\frac{(N+\alpha)(N+\beta)(N+
\alpha+\beta)}{(2N+\alpha+\beta-1)(2N+\alpha+\beta+1)}}
\nonumber\\
&&{}+\frac{2}{2N+\alpha+\beta}\sqrt{\frac{(N+\alpha)(N+\beta)(N+
\alpha+\beta)}{(2N+\alpha+\beta-1)(2N+\alpha+\beta+1)}}\,\,
a^\dagger + 
\frac{\beta^2-\alpha^2}{(2N+\alpha+\beta)(2N+\alpha+\beta+2)
}\,. 
\end{eqnarray}
Furthermore, making use of (4.15), (5.31), (5.32) and (5.33) 
we get
\begin{eqnarray}
&&\hat \pi = 
-a\,\frac{2N+\alpha+\beta-2}{2N+\alpha+\beta}\sqrt{\frac{(N+
\alpha)(N+\beta)(N+
\alpha+\beta)}{(2N+\alpha+\beta-1)(2N+\alpha+\beta+1)}}
\nonumber\\
&&{}+\frac{2N+\alpha+\beta+2}{2N+\alpha+\beta}\sqrt{\frac{(N+
\alpha)(N+\beta)(N+
\alpha+\beta)}{(2N+\alpha+\beta-1)(2N+\alpha+\beta+1)}}\,\,
a^\dagger + 
\frac{\beta^2-\alpha^2}{(2N+\alpha+\beta)(2N+\alpha+\beta+2)
}\,.\nonumber\\
&& 
\end{eqnarray}
Finally, by virtue of (4.22) and (5.32) we have
\begin{equation}
\hat k = i\left(-a\,\sqrt{\frac{(N+\alpha)(N+\beta)(N+
\alpha+\beta)}{(2N+\alpha+\beta-1)(2N+\alpha+\beta+1)}} + 
\sqrt{\frac{(N+\alpha)(N+\beta)(N+
\alpha+\beta)}{(2N+\alpha+\beta-1)(2N+\alpha+\beta+1)}}\,\,
a^\dagger\right)\,.
 \end{equation}
Consider now the system (4.44).  Using (4.46) and (5.33) we 
find that it can be reduced to the linear evolution equation 
in Hilbert space of the form (4.45), with
\begin{equation}
M = \sum_{i=1}^{k}\hat \pi_i\frac{F_i(\hat {\bf x})}{1-\hat 
x_i^2}\,,
\end{equation}
where in view of (4.43) and (4.47) $\hat x_i$ and $\hat 
\pi_i$ can be obtained immediately from (5.34) and (5.35) by 
formal replacement of $a$ and $N$ by $a_i$ and $N_i$, 
respectively.  That is $\hat x_i=\hat x(a=a_i,N=N_i)$ and 
$\hat \pi_i=\hat \pi(a=a_i,N=N_i)$.

The system (4.44) can be furthermore brought down to the 
Schr\"odinger equation (4.56).  On taking into account 
(5.36) we arrive at the following form of the corresponding 
Hamiltonian (4.53):
\begin{equation}
H = \frac{1}{2}\sum_{i=1}^{k}\left(\hat k_i\frac{F_i(\hat 
{\bf x})}{1-\hat x_i^2} + \frac{F_i(\hat {\bf x})}{1-\hat 
x_i^2}\hat k_i\right)\,,
\end{equation}
where $\hat k_i=\hat k(a=a_i,N=N_i)$.

We now discuss the Hilbert space counterpart (4.59) of 
the system (4.44).  Using (4.32) we get
\begin{eqnarray}
&&\hat \pi' = 
-a\,\frac{2(N-1)}{2N+\alpha+\beta}\sqrt{\frac{(N+
\alpha)(N+\beta)(N+
\alpha+\beta)}{(2N+\alpha+\beta-1)(2N+\alpha+\beta+1)}}
\nonumber\\
&&{}+\frac{2(N+\alpha+\beta+1)}{2N+\alpha+\beta}\sqrt{\frac{(N+
\alpha)(N+\beta)(N+
\alpha+\beta)}{(2N+\alpha+\beta-1)(2N+\alpha+\beta+1)}}\,\,
a^\dagger + 
\frac{2(\alpha-\beta)N(N+\alpha+\beta+1)}{(2N+\alpha+\beta)(2N+\alpha+\beta+2)
}\,.\nonumber \\
&&
\end{eqnarray}
Therefore, the operator $M'$ given by (4.60) takes the form
\begin{equation}
M' = \sum_{i=1}^{k}\hat \pi_i'\frac{F_i(\hat {\bf x})}{1-\hat 
x_i^2}\,,
\end{equation}
where $\hat \pi_i'=\hat \pi'(a=a_i,N=N_i)$.
We finally recall that Gegenbauer polynomials, Legendre polynomials
and Chebyshev polynomials are special cases of Jacobi polynomials.
We also point out that the actual treatment cannot be applied
in the case with Chebyshev polynomials of first kind 
$T_n(x)$.  Indeed, we have $T_{-1}(x)=x\ne0$, which 
implies violating of the recurrence formula (4.2) for $n=0$.
\subsection{Laguerre polynomials}
We finally study the case of (generalized) Laguerre 
polynomials $L_n^\alpha(x)$ within the actual approach.  
These polynomials correspond to
\begin{equation}
a=0,\qquad b=\infty,\qquad w(x)=x^\alpha e^{-x},
\end{equation}
where $\alpha>-1$.  Let $p_n$ be the normalized Laguerre 
polynomials, that is
\begin{equation}
p_n(x) = 
\left[\frac{\Gamma(n+1)}{\Gamma(n+\alpha+1)}\right]
^\frac{1}{2}L_n^\alpha(x).
\end{equation}
The coefficients in the recurrence formula (4.2) are
\begin{equation}
A_n=-\frac{1}{\sqrt{(n+1)(n+\alpha+1)}},\qquad 
B_n=\frac{2n+\alpha+1}{\sqrt{(n+1)(n+\alpha+1)}},\qquad 
C_n=\sqrt{\frac{n(n+\alpha)}{(n+1)(n+\alpha+1)}}\,.
\end{equation}
The constants $q_n$ and $r_n$ in (4.3) are of the form
\begin{equation}
q_n = 
\frac{(-1)^n}{\sqrt{\Gamma(n+1)\Gamma(n+\alpha+1)}},\qquad 
r_n = -n(n+\alpha).
\end{equation}
The expressions for the polynomial $X$ and the constant 
$K_n$ in the Rodrigues formula are
\begin{equation}
X=x,\qquad K_n = \sqrt{\Gamma(n+1)\Gamma(n+\alpha+1)}\,.
\end{equation}
Now, owing to (4.11) and (5.43) we get
\begin{equation}
\hat x = -a\sqrt{N+\alpha} - \sqrt{N+\alpha}\,\,a^\dagger + 2N 
+ \alpha + 1.
\end{equation}
Furthermore, making use of (4.15) one obtains
\begin{equation}
\hat \pi = \hbox{$\frac{1}{2}$}(a\sqrt{N+\alpha} - 
\sqrt{N+\alpha}\,\,a^\dagger - 1).
\end{equation}
Hence, in view of (4.22)
\begin{equation}
\hat k = \hbox{$\frac{i}{2}$}(a\sqrt{N+\alpha} - 
\sqrt{N+\alpha}\,\,a^\dagger).
\end{equation}
We are now in a position to write down eqs.\ (4.46) and 
(4.53) in the case with Laguerre polynomials.  We have
\begin{equation}
M = \sum_{i=1}^{k}\hat \pi_i\frac{F_i(\hat {\bf x})}{\hat 
x_i}
\end{equation}
and
\begin{equation}
H = \frac{1}{2}\sum_{i=1}^{k}\left(\hat k_i\frac{F_i(\hat 
{\bf x})}{\hat x_i} + \frac{F_i(\hat {\bf x})}{\hat x_i}\hat 
k_i\right)\,,
\end{equation}
where $\hat x_i=\hat x(a=a_i,N=N_i)$, $\hat \pi_i=\hat 
\pi(a=a_i,N=N_i)$ and $\hat k_i=\hat k(a=a_i,N=N_i)$.  We 
now return to (4.30).  Using (4.32) we find
\begin{equation}
\hat \pi' = -\sqrt{N+\alpha}\,\,a^\dagger + N.
\end{equation}
Therefore, the operator (4.60) is
\begin{equation}
M' = \sum_{i=1}^{k}\hat \pi_i'\frac{F_i(\hat {\bf x})}{\hat 
x_i},
\end{equation}
where $\hat \pi_i'=\hat \pi'(a=a_i,N=N_i)$.
\section{Conclusion}
We have introduced in this work the approach generalizing 
the classical methods for linearization of nonlinear 
dynamical systems such as the Carleman embedding technique 
and Koopman formalism.  Moreover, in the light of the 
observations of section 5 the Hilbert space approach 
developed by the author can be also regarded as a special 
case of the treatment introduced herein.  The role played by 
classical orthogonal polynomials in the presented formalism 
is remarkable.  On the one hand, the classical orthogonal 
polynomials have been shown to be the most natural tool 
for the Hilbert space linearization of nonlinear dynamical 
systems.  On the other hand, the stability of classical 
orthogonal polynomials with respect to the nonlinear 
time-evolution has been demonstrated to be one of the 
properties which actually determine these polynomials via
``canonical'' algebraic relations like (4.16) and (4.21).  
As a consequence of the general algorithm the new methods 
have been found in this work for linearization of nonlinear 
dynamical systems connected with an arbitrary classical 
orthogonal polynomial, excluding the case with Hermite 
polynomials which has been shown herein to correspond to the 
classical Koopman approach.  We note that in opposite to the 
existing approaches mentioned above, these methods cover the 
case of systems with the phase space different from the 
whole ${\bf R}^k$.  For example, in the case when the 
Hilbert space counterpart of the system (4.44) is (4.45) 
with $M$ given by (5.37), we have the restrictive condition 
$x_i\ne1$, $i=1$, $\ldots\,$,~$k$, where {\bf x} satisfies 
(4.44).  Analogously, (5.49) leads to the requirement that 
$x_i\ne0$, $i=1$, $\ldots\,$,~$k$, where {\bf x} fulfils 
(4.44).  It is suggested that the phase space of the 
linearized 
nonlinear dynamical system (4.44) should coincide with a 
subset of $(a,b)^k$ (see (4.1) and (4.38)), where $(a,b)$ is the 
interval associated with the corresponding classical 
orthogonal polynomial.  Therefore, the case of (5.37) refers to
the system (4.44) such that the solution remains in a 
$k$-dimensional cube $(-1,1)^k$.  Clearly, an arbitrary 
system performing finite motion can be reduced to such one 
by appropriate rescaling.  On the other hand, (5.49) 
corresponds to systems (4.44) with ${\bf x}\in{\bf 
R}_+^k$, where ${\bf R}_+$ is the set of positive real 
numbers.  The well-known example of such systems are rate 
equations of chemical kinetics.  The experience with the 
Koopman linearization, the Carleman embedding technique and 
the Hilbert space approach indicates that the formalism 
introduced herein would be a useful tool in the theory of 
nonlinear dynamical systems.  Last but not least, we point 
out that results of this paper would be of importance in the 
theory of orthogonal polynomials.  Indeed, it might be 
observed that the formulas (4.8) and (4.9) as well as the 
relations (4.16) and (4.21) mentioned above seem to provide 
a new algebraic approach to the theory of classical 
orthogonal polynomials.  We also point out that relations (4.51)
amount a generalization of the Heisenberg algebra (C.1).  On the
other hand, we have shown earlier (see section 4) that this algebra
can be interpreted as a condition for stability of solutions to (3.2).
\begin{acknowledgements}
I would like to thank a referee for helpful comments.
\end{acknowledgements}
\renewcommand{\theequation}{\Alph{section}.\arabic{equation}}
\appendix
\section{Bose operators and occupation number 
representation}
We recall the basic properties of the standard Bose 
operators and the occupation number representation.  The 
Bose creation (${\bf a}^\dagger $) and annihilation (${\bf 
a}$) operators, where ${\bf a}^\dagger =(a_1^\dagger ,\ldots 
,a_k^\dagger )$ and ${\bf a}=(a_1,\ldots ,a_k)$, satisfy the 
Heisenberg-Weyl algebra
\begin{equation}
[a_i,a_j^\dagger ]=\delta_{ij},\qquad [a_i,a_j]=[a_i^\dagger 
,a_j^\dagger ]=0,\qquad i,\,j=1,\,\ldots \,,\,k.
\end{equation}
The Hermitian operators $N_i=a_i^\dagger a_i$, $i=1$, 
$\ldots\,$,~$k$, are called the number operators.  These 
operators obey
\begin{equation}
[N_i,N_j]=0,\qquad [N_i,a_j]=-\delta_{ij}a_i,\qquad 
[N_i,a_j^\dagger ]=\delta_{ij}a_i^\dagger ,\qquad i,\,j=1,\,\ldots \,,\,k.
\end{equation}
Let us assume that there exists in the Hilbert space of 
states $\cal H$, a unique, normalized vector $|{\bf 
0}\rangle$ (vacuum vector) such that
\begin{equation}
{\bf a}|{\bf 0}\rangle = {\bf 0}.
\end{equation}
We also assume that there is no nontrivial closed subspace 
of $\cal H$ which is invariant under the action of the Bose 
operators.  The state vectors $|{\bf n}\rangle$, ${\bf 
n}\in{\bf Z}_+^k$, defined as
\begin{equation}
|{\bf n}\rangle = \left(\prod_{i=1}^{k}\frac{a_i^\dagger }
{\sqrt{n_i!}}\right)|{\bf 0}\rangle,
\end{equation}
are the common eigenvectors of the number operators, i.e.
\begin{equation}
{\bf N}|{\bf n}\rangle = {\bf n}|{\bf n}\rangle.
\end{equation}
These vectors form the orthonormal basis of $\cal H$, that 
is
\begin{eqnarray}
 &&\langle {\bf n}|{\bf m}\rangle = 
\prod_{i=1}^{k}\delta_{n_im_i},\\
 &&\sum_{{\bf n}\in{\bf Z}_+^k}|{\bf n}\rangle\langle {\bf 
n}| = I.
\end{eqnarray}
The action of the Bose operators on the vectors $|{\bf 
n}\rangle$ has the following form:
\begin{equation}
a_i|{\bf n}\rangle=\sqrt{n_i}\,|{\bf n}-{\bf 
e}_i\rangle,\qquad a_i^\dagger |{\bf 
n}\rangle=\sqrt{n_i+1}\,|{\bf 
n}+{\bf e}_i\rangle,\qquad i=1,\,\ldots \,,\,k.
\end{equation}
We finally write down the following formulas corresponding 
to the case with $k=1$, which were frequently used 
throughout this work:
\begin{equation}
f(N)a=af(N-1),\qquad a^\dagger f(N)=f(N-1)a^\dagger ,
\end{equation}
where $N=a^\dagger a$ is the number operator,
\begin{equation}
a|n\rangle=\sqrt{n}\,|n-1\rangle,\qquad a^\dagger 
|n\rangle=\sqrt{n+1}\,|n+1\rangle.
\end{equation}
\section{Coherent states}
We now outline the main facts about the standard coherent 
states.  Consider the coherent states $|{\bf z}\rangle$, 
where ${\bf z}\in{\bf C}^k$, that is the eigenvectors of the 
Bose annihilation operators
\begin{equation}
{\bf a}|{\bf z}\rangle = {\bf z}|{\bf z}\rangle.
\end{equation}
The normalized coherent states can be defined as
\begin{equation}
|{\bf z}\rangle = 
\exp\left(-\frac{1}{2}\sum_{i=1}^{k}|z_i|^2\right)\exp\left(
\sum_{i=1}^{k}z_ia_i
^\dagger \right)|{\bf 0}\rangle,
\end{equation}
where $|{\bf 0}\rangle$ is the vacuum vector.  The coherent 
states are not orthogonal.  We have
\begin{equation}
\langle {\bf z}|{\bf w}\rangle = 
\exp\left[-\frac{1}{2}\sum_{i=1}^{k}(|z_i|^2+|w_i|^2-2z_i^*
w_i)\right].
\end{equation}
These states form the complete (overcomplete) set.  Namely,
\begin{equation}
\int_{{\bf R}^{2k}}d\mu({\bf z})\,|{\bf z}\rangle\langle 
{\bf z}| = I,
\end{equation}
where
\begin{equation}
d\mu({\bf z}) = \prod_{i=1}^{k}\frac{1}{\pi}d({\rm 
Re}z_i)\,d({\rm Im}z_i).
\end{equation}
The passage from the occupation number representation to the 
coherent state representation is given by the kernel
\begin{equation}
\langle {\bf n}|{\bf z}\rangle = 
\left(\prod_{i=1}^{k}\frac{z_i^{n_i}}{\sqrt{n_i!}}\right)
\exp\left(-\frac{1}{2}\sum_{i=1}^{k}|z_i|^2\right)\,.
\end{equation}
On taking into account (B.4), (A.7) and (B.6) we arrive at 
the Fock-Bargmann space of analytic (entire) functions 
specified by the inner product
\begin{equation}
\langle \phi|\psi\rangle = \int_{{\bf R}^{2k}}d\mu({\bf 
z})\,\exp\left(-\sum_{i=1}^{k}|z_i|^2\right)\left(\tilde 
\phi({\bf z}^*)\right)^*\tilde \psi({\bf z}^*),
\end{equation}
where $\tilde \phi({\bf z}^*)=\langle {\bf 
z}|\phi\rangle\exp\left(\frac{1}{2}\sum_{i=1}^{k}|z_i|^2
\right)$ and ${\bf z}^*=(z_1^*,\ldots ,z_k^*)$.  We remark that in
view of (B.4) and (B.6) the polynomials
\begin{equation}
p_{\bf n}({\bf z}) = \prod_{i=1}^{k}\frac{z_i^{n_i}}{\sqrt{n_i!}}
\end{equation}
form the orthonormal complete set in the Fock-Bargmann space
that is, we have
\begin{equation}
\int_{{\bf R}^2}d\mu({\bf
z})\,\exp\left(-\sum_{i=1}^{k}|z_i|^2\right)(p_{\bf n}({\bf
z}^*))^*p_{\bf m}({\bf z}^*) = \delta_{\bf nm},
\end{equation}
where $\delta_{\bf nm}=\prod_{i=1}^{k}\delta_{n_im_i}$. 
\section{Position and momentum operators}
We finally collect some basic properties of position and 
momentum operators.  The position ($\hat {\bf q}$) and 
momentum ($\hat {\bf p}$) operators, where $\hat {\bf 
q}=(\hat q_1,\ldots ,\hat q_k)$ and $\hat {\bf p}=(\hat 
p_1,\ldots ,\hat p_k)$ satisfy the Heisenberg algebra
\begin{equation}
[\hat q_r,\hat p_s]=i\delta_{rs},\qquad [\hat q_r,\hat 
q_s]=[\hat p_r,\hat p_s]=0,\qquad r,\,s=1,\,\ldots \,,\,k.
\end{equation}
These operators are related to the Bose operators by
\begin{eqnarray}
\hat {\bf q} &=& \hbox{$\frac{1}{\sqrt{2}}$}({\bf a}+{\bf a}^\dagger 
),\qquad \hat {\bf p}=\hbox{$\frac{i}{\sqrt{2}}$}({\bf 
a}^\dagger - {\bf a}),\nonumber\\
{\bf a} &=& \hbox{$\frac{1}{\sqrt{2}}$}(\hat {\bf q}+i\hat 
{\bf p}),\qquad {\bf a}^\dagger = 
\hbox{$\frac{1}{\sqrt{2}}$}(\hat {\bf q}-i\hat {\bf p}).
\end{eqnarray}
An immediate consequence of (C.1) is
\begin{equation}
[\hat {\bf p},f(\hat {\bf q})] = -i\frac{\partial f(\hat 
{\bf q})}{\partial \hat {\bf q}}.
\end{equation}
Consider now the eigenvectors $|{\bf q}\rangle$, ${\bf 
q}\in{\bf R}^k$, of the position operators
\begin{equation}
\hat {\bf q}|{\bf q}\rangle = {\bf q}|{\bf q}\rangle.
\end{equation}
The normalized eigenvectors can be expressed by
\begin{equation}
|{\bf q}\rangle = 
\pi^{-\frac{k}{4}}\exp(\hbox{$\frac{1}{2}$}{\bf 
q}^2)\exp[-\hbox{$\frac{1}{2}$}({\bf a}^\dagger - 
\sqrt{2}\,{\bf q})^2]|{\bf 0}\rangle,
\end{equation}
where $|{\bf 0}\rangle$ is the vacuum vector.  The states 
$|{\bf q}\rangle$ form the orthogonal and 
complete set, namely
\begin{eqnarray}
 && \langle {\bf q}|{\bf q}'\rangle = \delta({\bf q}-{\bf 
q}'),\\
 &&\int d^kq\,|{\bf q}\rangle\langle {\bf q}| = I.
\end{eqnarray}
The passage from the coordinate representation spanned by 
the vectors $|{\bf q}\rangle$ to the occupation number 
representation is given by the kernel
\begin{equation}
\langle {\bf n}|{\bf q}\rangle = 
\left(\prod_{i=1}^{k}p_{n_i}(q_i)\right)\exp(-\hbox{$\frac{1}
{2}$}{\bf q}^2),
\end{equation}
where $p_n(q)$ are normalized Hermite polynomials, that is 
\begin{equation}
p_n(q)=\pi^{-\frac{1}{4}}2^{-\frac{n}{2}}\frac{1}{\sqrt{n!}
}H_n(q).
\end{equation}

\end{document}